\documentclass[preprint]{aastex}  

\shorttitle{Yamamura, Kawaguchi, \& Ridgway}
\shortauthors{Identification of SH in R And}

\begin{document}

\title{Identification of SH $\Delta v=1$ ro-vibrational lines in R And}

\author{I. Yamamura}
\affil{Astronomical Institute `Anton Pannekoek', University of Amsterdam, \\
      Kruislaan 403, 1098 SJ, Amsterdam, the Netherlands}
\email{yamamura@astro.uva.nl}

\author{K. Kawaguchi}
\affil{Faculty of Science, Okayama University, \\
3-1-1, Tsushima-naka, Okayama 700-8530, Japan}
\email{okakent@cc.okayama-u.ac.jp}

\and

\author{S. T. Ridgway}
\affil{National Optical Astronomy Observatories\footnotemark[1] \\
P.O. Box 26732, Tucson, AZ 85726, U.S.A.}
\email{ridgway@noao.edu}

\footnotetext[1]{
   Operated by the Association of Universities
   for Research in Astronomy under cooperative agreement
   with the National Science Foundation.
}

\begin{abstract}
We report the identification of SH $\Delta v=1$ ro-vibrational lines in
the published high-resolution infrared spectrum of the S-type star, R And.
This is the first astronomical detection of this molecule.
The lines show inverse P-Cygni profiles, indicating infall motion
of the molecular layer due to stellar pulsation.
A simple spherical shell model with a constant infall velocity is adopted
to determine the condition of the layer.
It is found that a single excitation temperature of 2200 K
reproduces the observed line intensities satisfactory.
SH is located in a layer from 1.0 to $\sim 1.1$ stellar radii,
which is moving inward with a velocity of 9 km\,s$^{-1}$.
These results are consistent with the previous measurements of
CO $\Delta v=3$ transitions.
The estimated molecular abundance SH/H is $1\times 10^{-7}$,
consistent with a thermal equilibrium calculation.

\end{abstract}

\keywords{infrared: stars---line: identification---stars: AGB and post-AGB
---stars:atmospheres---stars: individual(R~And)}

\section{Introduction}

Chemistry of sulfur-containing species in space is especially interesting
due to its chemical activity and relatively high abundance \citep{Duley80}.
It is known that abundances of sulfur-bearing molecules are strongly
influenced by the presence of shocks \citep{Hartquist80}.
So far, 14 sulfur-bearing species have been identified in space;
CS, SO, NS, SiS, SO$^+$, SH$_2$, OCS, CCS, C$_3$S, H$_2$CS, HNCS,
SO$_2$, HCS$^+$, and CH$_3$SH.
However, the simplest sulfur compound, SH, has not been
detected in spite of radioastronomical searches using the $\Lambda$-type
doubling transitions \citep{Meeks69,Heiles71}.

While sulfur chemistry in the circumstellar envelope of red giant stars
has been studied extensively \citep[e.g.,][]{Omont93},
that in the atmosphere has been regarded
as in the thermal equilibrium state \citep{Tsuji64,Tsuji73}.
In the atmosphere of oxygen-rich giants,
SH is the molecule first formed from the sulfur atom,
at the temperature of about 2000 K.
The SH abundance decreases when temperature is below $\sim 1500$ K,
and H$_2$S and SiS become the dominant species.
On the other hand, \citet{Yamamura99a} report the detection of
the $\nu_3$ infrared band of SO$_2$ in the spectra of
oxygen-rich Mira variables obtained by the Short-Wavelength Spectrometer
\citep[SWS:][]{deGraauw96} on board the Infrared Space Observatory
\citep[ISO:][]{Kessler96}.
Their model analysis indicates that the molecules are located
in the extended atmosphere, at about five stellar radii.
The abundance of SO$_2$ is 5--10 orders of magnitude larger than
the values in thermal equilibrium \citep{Tsuji73,Woitke99}.
The result strongly suggests the presence of non-equilibrium processes
in the extended atmosphere, probably related to shocks due to
stellar pulsation \citep{Beck92,Duari99}.

In this paper, we report the detection of SH ro-vibrational transition
lines in the published high-resolution infrared spectrum of
the S-type star, R And.
We discuss the physical and chemical conditions of the SH layer in the star.

\section{Identification of SH lines}

The data used in this study were obtained with the Fourier-transform
spectrometer at the Kitt Peak National Observatory 4\,m telescope,
and were published in \citet{Ridgway84}.
R And is a Mira variable with a period of 409 days
and a visual amplitude of 9.1 mag \citep{Kholopov88}.
The spectral type ranges in S3,5e--S8,8e.
The observation of R And was performed at the optical variable phase
of $\phi=0.82$.
Unfortunately, the reduced data were lost after several updates of
archive formats at the observatory.
Therefore, we re-digitized the data from the printed spectrum.
The results may be as accurate as the original within
$\sim 0.02$ cm$^{-1}$ in wavenumber and a few \% in intensity.
The spectrum was shifted by 5.2 km\,s$^{-1}$ to longer wavelengths
to adjust the absorption minimum of HCl lines at their rest frequencies
\citep{Ridgway84}. No correction for terrestrial motion was applied.
In their original paper, \citet{Ridgway84} give identifications
of OH, NH, CH, SiO, CS, HCl, and atomic lines.
The spectrum of R And, especially in the wavelengths shorter
than the SiO first-overtone bandheads, is dominated by HCl lines.
OH and NH lines are detected, but much less prominently than
in the oxygen-rich stars like $\alpha$ Ori and $\alpha$ Tau.
There are many strong lines left unidentified.
Figure~\ref{fig1} shows the spectrum of R And between 2700 and 2750 cm$^{-1}$.
The positions of SH and HCl ro-vibrational transitions are indicated.
The frequencies are calculated from the molecular constants
given by \citet{Ram95} for SH, and are taken from HITRAN database
1996 edition \citep{Rothman92} for HCl.
It is obvious that many unidentified strong lines are
attributed to SH transitions.
We assigned 39, 41, and 11 transitions in the $v=1-0$, $2-1$,
and $3-2$ bands, respectively, between 2500 and 2778 cm$^{-1}$.
The lower state energies of these transitions are up to
$7000$ cm$^{-1}$, indicating that the molecules are highly excited.
The isotope $^{34}$SH may also be detected. The frequencies of $^{34}$SH
lines are estimated from the molecular constants of SH using the relation of
reduced mass \citep{Herzberg50}.

\section{Modeling}

We analyze the observed SH line profiles with a simple model.
The SH (and also HCl) lines in R And show inverse P-Cygni profiles,
indicating that the molecular layer is moving inward to the star
due to stellar pulsation.
We, therefore, apply a spherical shell model with a constant infall velocity.
The molecules in front of the star cause absorption,
while those extended in the blank sky contribute as emission.
Considering that the shell may be in the atmosphere,
we adopt an exponential density law, $n(r) \propto \exp(-r/r_0)$,
where $r_0$ is the scale height.
The star is assumed to be a 3000 K blackbody.
For simplification, a constant excitation temperature is adopted, and
the energy population of the molecule is calculated by assuming
local thermodynamic equilibrium (LTE). This may be
justified if the molecular shell is thin and in the high density region
near the photosphere.
The line intensity of SH is calculated based on \citet{Benidar91},
which takes account of the Herman-Wallis effect.
The spectrum is normalized by the stellar continuum.
No smoothing is applied.

The excitation temperature can be determined from the relative
intensities of the lines at different energy levels;
higher temperatures excite the molecules to higher energy levels
and increase the line strength from these levels.
We find that 2200 K is most reasonable for the present case.
A turbulent velocity of 6 km\,s$^{-1}$ and
infall velocity of $9$ km\,s$^{-1}$ reproduce the line profiles.
The model spectrum is shifted by $-14$ km\,s$^{-1}$.
Considering the terrestrial velocity of $-2.7$ km\,s$^{-1}$
at transit of the observation, and the shift applied
by \citet{Ridgway84}, $+5.2$ km\,s$^{-1}$, a radial velocity of
$-16.5$ km\,s$^{-1}$ is obtained.
The inner radius and the scale height of the SH shell
is 1.0 and 0.08 times the stellar radius, respectively.
The SH column density in the shell is $4.0 \times 10^{20}$ cm$^{-2}$.
The uncertainties of the parameters, which changes the
model spectrum by about 10 \%, are $\pm 100$ K for excitation temperature,
a factor of two for column density, $\pm 0.01$ R$_\star$ for
scale hight, and $\pm 1$ km\,s$^{-1}$ for the velocities.
The radial velocity has an error of $\pm 2$ km\,s$^{-1}$
due to the re-digitization process.
The model spectrum is compared with the observation in Figure~\ref{fig2}.
The wavelength regions are selected so that
the contamination of other spectral lines is minimal, and that
a wide range of energy levels is covered.
Strong lines in Figure~\ref{fig2} are listed in Table~\ref{tbl1}.
The fit is satisfactory in most of the lines.
Fitting of $^{34}$SH lines results in an isotopic abundance ratio
of 5--10 \%.
We note that the same parameters also give reasonable fits
for the HCl lines with a column density of $3.0 \times 10^{19}$ cm$^{-2}$.

\section{Discussion}

Why are SH lines so prominent in R And, an S-type star?
S-type stars are characterized by their chemical anomaly
of similar carbon and oxygen abundances.
This leaves few C and O atoms for further chemical processes,
after formation of CO molecules at 4--5000 K.
SH should be quite abundant in the atmosphere of oxygen-rich stars,
but a minor product in a carbon-rich environment \citep{Tsuji73}.
Since we do see SiO lines, but no CS line, in the spectrum of R And,
the star is slightly oxygen-rich.
Nevertheless, it is not expected that SH is much more abundant
in S-type stars than O-rich stars.
This also holds for HCl, which is the most abundant Cl-bearing molecule
in an oxygen-rich environment, although the absolute abundance is
by one order of magnitude lower than SH.
\citet{Ridgway84} suggested that the atmosphere of S-type stars
is more transparent than O-rich stars, so that the lines of minor species
could be stronger due to larger path length.
We especially emphasize the contribution of the H$_2$O molecules.
In the same paper, Ridgway et al. reported that an enormous number of
lines heavily blanket the 2400--2800 cm$^{-1}$ region of
the $o$ Cet spectrum.
They suspected that these are highly excited water lines.
This is supported by the ISO/SWS observation of $o$ Cet \citep{Yamamura99b}.
Despite the relatively poor resolution of the SWS spectrum,
they demonstrate that the star shows highly excited H$_2$O lines
in the 3.5--4.0 $\mu$m region, arising from hot (2000 K)
and optically thick molecular layer.
Probably, in O-rich stars, SH (and HCl) lines are weaker because of
the shorter path length, and also heavily contaminated by the hot water lines.
On the other hand, H$_2$O molecules are not favored in the atmosphere
of S-type stars. The thermal equilibrium abundance
of H$_2$O in S-type stars is 3--4 orders lower than in O-rich stars
\citep{Tsuji64}.

The derived infall velocity and the excitation temperature of the SH layer
are consistent with the measurement of CO $\Delta v=3$ lines
by \citet{Hinkle84} at $\phi=0.80$, indicating that the SH molecules
are in the same region as hot CO near the photosphere.
Comparison with the CO column density, $2.6\times 10^{24}$ cm$^{-2}$,
given by \citet{Hinkle84}, leads to SH/H~$=1\times 10^{-7}$.
In thermal equilibrium at the temperature of 2200 K and
gas pressure of $\log(P_g)=1.0$, the SH abundance is
$2\times 10^{-7}$ \citep{Tsuji73,Tsuji99}.
This value is insensitive to the C/O ratio as long as C/O~$< 1$.
The present estimate of SH abundance in R And, although it is rather crude,
is consistent with this calculated value.
The sulfur chemistry in SH layer still follows thermal equilibrium,
because of its high density and high temperature.

We see no clear evidence of distinct velocity and/or
different temperature component of SH lines in the spectrum.
This implies that the SH molecules are distributed only in
a thin layer in the atmosphere.
In thermal equilibrium, SH is most abundant around 1800 K,
and then is rather quickly transformed to H$_2$S or SiS below $\sim 1500$ K
\citep{Tsuji73}.
Otherwise, non-equilibrium chemical reactions may lead to completely
different compositions in the extended atmosphere,
e.g. the enhancement of SO$_2$ in oxygen-rich stars
\citep{Yamamura99a,Beck92}.
We could not find any clear indications of H$_2$S or SO$_2$
in the spectrum of R And.
The upper limit of the H$_2$S column density is at least a factor of 10 larger
than the observed amount of SH.
A dioxide molecule SO$_2$ may not be abundant
in the atmospheres of S-type stars.
Other possible candidates, SO or SiS, have no transition in
the present wavelength coverage.

\acknowledgments
The authors are grateful to Prof. T. Tsuji for his suggestion
on thermal equilibrium chemistry and calculations of SH abundance.
I.Y. acknowledges financial support from a NWO PIONIER grant.

\clearpage

\figcaption[fig1a.ps]{
A part of the spectrum of R~And obtained by \citet{Ridgway84}
is presented. The positions of $\Delta v = 1$ transitions of
SH and $^{34}$SH (upper), and HCl (lower) are indicated by ticks.
The notation of the transitions is the electronic angular momentum,
the upper vibrational level, and the rotational line index for SH,
and the rotational line index,
the isotope index (5: H$^{35}$Cl, 7: H$^{37}$Cl), and the upper vibrational
level for HCl, respectively.
\label{fig1}
}

\figcaption[fig2.ps]{
The synthesized model spectrum (upper) is compared with the observation
in six wavelength regions. The regions are chosen so that contamination
of other spectral lines is minimum, and that
a wide range of energy levels is covered.
The model assumes a spherical shell falling toward the
stellar surface with a velocity of 9 km\,s$^{-1}$.
The molecules are distributed from 1.0 R$_\star$ with a scale height
of 0.08 R$_\star$.
The model assumes constant excitation temperature of 2200 K,
column density of $4.0\times 10^{20}$ cm$^{-2}$,
and Doppler broadening velocity of 6 km\,s$^{-1}$.
The model spectrum is shifted by $-14$ km\,s$^{-1}$.
The lines indicated by the ticks are listed in Table~\ref{tbl1}.
\label{fig2}
}

\begin{figure}
   \figurenum{1}
   \epsscale{0.5}
   
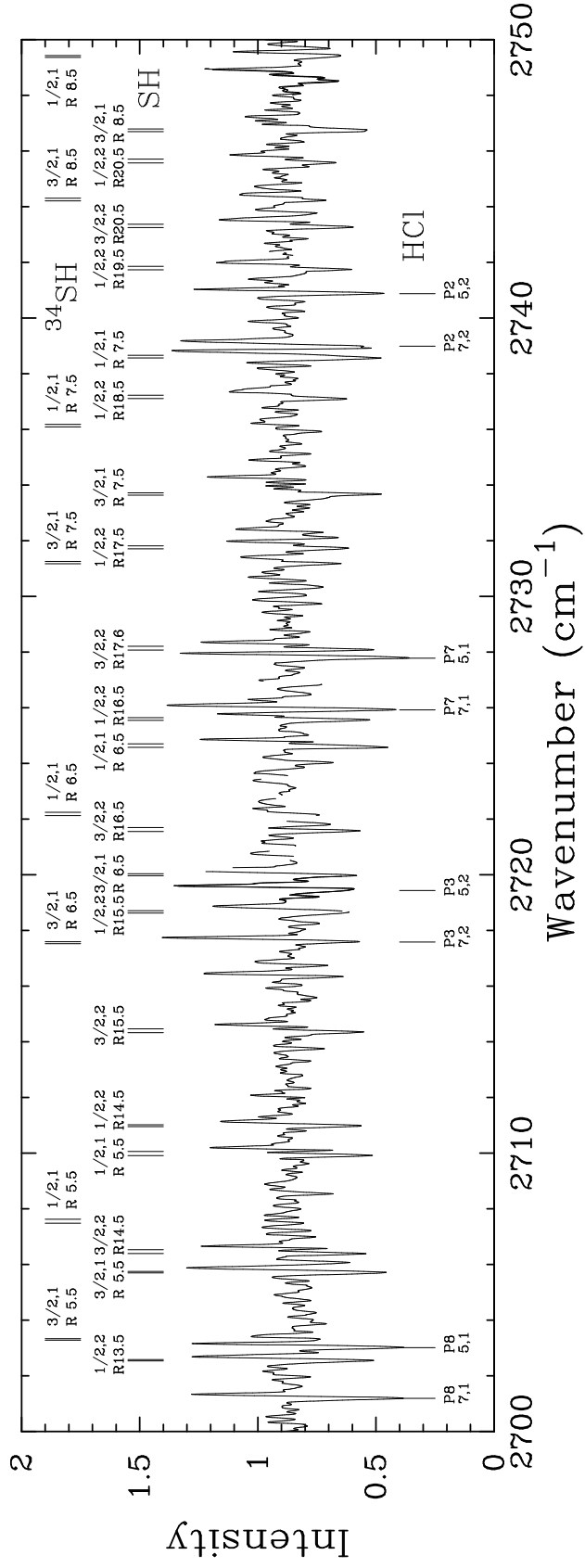
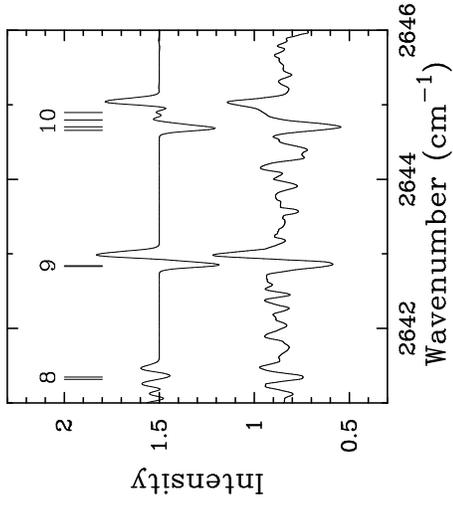
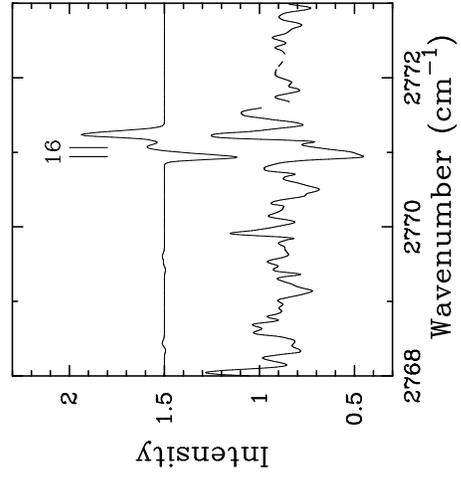
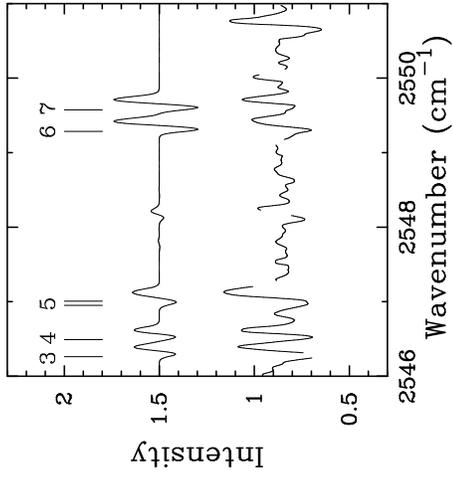
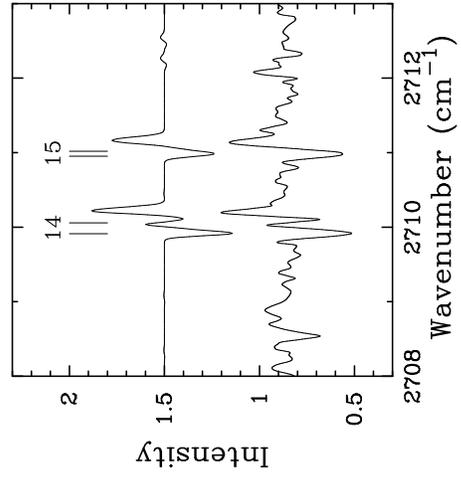
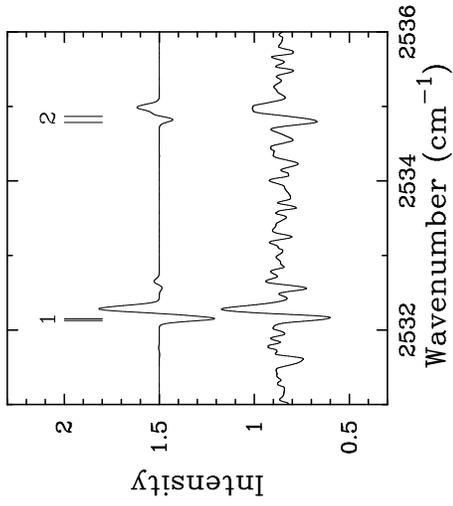
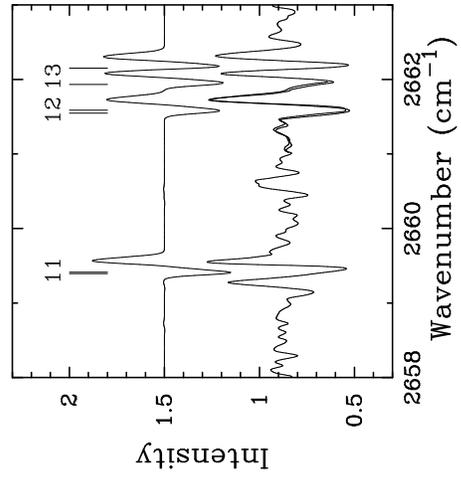
\plotone{fig1a.ps}
\end{figure}

\begin{figure}
   \figurenum{2}
   \epsscale{0.8}
   \plotone{fig2.ps}
\end{figure}

\clearpage

\begin{deluxetable}{rlllr}
\footnotesize
\tablecaption{SH lines indicated in Figure~\ref{fig2}.\label{tbl1}}
\tablewidth{0pt}
\tablehead{
\colhead{No.\tablenotemark{a}} &
\colhead{Frequency} &
\colhead{$\Omega$\tablenotemark{b}} &
\colhead{Transition\tablenotemark{c}} &
\colhead{$E_l$\tablenotemark{d}} \\
&
\colhead{cm$^{-1}$} &
&
&
\colhead{cm$^{-1}$}
}
\startdata
1  & 2532.1257 & 3/2 & $1-0$ P 3.5ff & $  110.7175$ \\
1  & 2532.1486 & 3/2 & $1-0$ P 3.5ee & $  110.6810$ \\
2  & 2534.7858 & 1/2 & $3-2$ R 7.5ee & $ 6038.9539$ \\
2  & 2534.8657 & 1/2 & $3-2$ R 7.5ff & $ 6040.7705$ \\
3  & 2546.2628 & 1/2 & $2-1$ R 1.5ee & $ 2994.8376$ \\
4  & 2546.4908 & 1/2 & $2-1$ R 1.5ff & $ 2995.3755$ \\
5  & 2546.9511 & 1/2 & $3-2$ R 8.5ee & $ 6192.1858$ \\
5  & 2547.0067 & 1/2 & $3-2$ R 8.5ff & $ 6194.1577$ \\
6  & 2549.2843 & 1/2 & $1-0$ P 2.5ff & $  446.0913$ \\
7  & 2549.5750 & 1/2 & $1-0$ P 2.5ee & $  445.2626$ \\
8  & 2641.3159 & 3/2 & $3-2$ R24.5ee & $10354.2802$ \\
8  & 2641.3458 & 3/2 & $3-2$ R24.5ff & $10358.7197$ \\
9  & 2642.8296 & 3/2 & $1-0$ R 1.5ee & $    0.0000$ \\
9  & 2642.8395 & 3/2 & $1-0$ R 1.5ff & $    0.0037$ \\
10 & 2644.6568 & 1/2 & $1-0$ R 1.5ee & $  396.9951$ \\
10 & 2644.7048 & 3/2 & $2-1$ R 8.5ee & $ 3286.0606$ \\
10 & 2644.7949 & 3/2 & $2-1$ R 8.5ff & $ 3286.4309$ \\
10 & 2644.8958 & 1/2 & $1-0$ R 1.5ff & $  397.5540$ \\
11 & 2659.3953 & 3/2 & $1-0$ R 2.5ee & $   46.1293$ \\
11 & 2659.4143 & 3/2 & $1-0$ R 2.5ff & $   46.1440$ \\
12 & 2661.5489 & 1/2 & $2-1$ R 9.5ee & $ 3889.9056$ \\
12 & 2661.5866 & 1/2 & $2-1$ R 9.5ff & $ 3892.0848$ \\
13 & 2661.9331 & 1/2 & $1-0$ R 2.5ee & $  445.2626$ \\
13 & 2662.1518 & 1/2 & $1-0$ R 2.5ff & $  446.0913$ \\
14 & 2709.9111 & 1/2 & $1-0$ R 5.5ee & $  705.7060$ \\
14 & 2710.0575 & 1/2 & $1-0$ R 5.5ff & $  707.2638$ \\
15 & 2710.9529 & 1/2 & $2-1$ R14.5ff & $ 5042.3624$ \\
15 & 2711.0189 & 1/2 & $2-1$ R14.5ee & $ 5039.8647$ \\
16 & 2770.9408 & 3/2 & $1-0$ R10.5ee & $ 1074.9261$ \\
16 & 2771.0636 & 3/2 & $1-0$ R10.5ff & $ 1075.6188$ \\
\enddata

\tablenotetext{a}{Numbers in Figure~\ref{fig2}.}
\tablenotetext{b}{Electronic angular momentum index.}
\tablenotetext{c}{Vibrational and rotational transition indices.}
\tablenotetext{d}{Lower state energy relative to $v=0,J=1.5,\Omega=3/2$.}

\end{deluxetable}

\end{document}